\documentclass[]{spie}  

\usepackage{lineno}

\usepackage{amsmath,amsfonts,amssymb}
\usepackage{graphicx}
\usepackage{xcolor}
\usepackage[colorlinks=true, allcolors=blue]{hyperref}

\title{Phase Retrieval using Nonlinear Curvature Sensing within Convergent Beams}

\author[a]{Justin R. Crepp}
\author[a]{Caleb G. Abbott}
\author[a]{James Smous}
\author[a]{Matthew Engstrom}
\author[a]{Brian Sands}

\affil[a]{University of Notre Dame, Department of Physics and Astronomy, Notre Dame, IN, USA}
\authorinfo{Send correspondence to jcrepp@nd.edu}

\begin{document}
\maketitle


\begin{abstract} 
{Path-length diversity methods may be used for adaptive optics (AO) systems to retrieve phase and amplitude information} by measuring intensity across multiple planes. Observations that rely on free-space propagation, such as the nonlinear curvature wavefront sensor (WFS), have been shown to offer excellent sensitivity and robustness to scintillation. However, the default design results in a large opto-mechanical footprint due to unavoidable geometric-optics and wave-optics effects. Measurements recorded in a convergent beam would improve instrument compactness, while concentrating light into smaller detector regions of interest, improving signal-to-noise ratio and possibly wavefront reconstruction speed. In this paper, we study path-length diversity wavefront sensing using four planes of contemporaneous intensity measurements made in a convergent beam. We develop a physical optics propagation model and validate the model by performing wavefront reconstructions in both simulations and lab experiments. The manuscript’s core contribution is a practical, intensity-domain, Fourier-transform-based recipe to use a conventional multi-plane Gerchberg-Saxton (or comparable) reconstruction pipeline with convergent-beam measurements, enabling a compact optical layout. We find that this approach offers practical benefits over an equivalent free-space wavefront sensor, in particular reducing size, weight, complexity and cost.  
\end{abstract} 




\section{Introduction}\label{sec:intro}

Diffraction and interference of spatially coherent radiation may be used to estimate the electric field of light waves that become aberrated by inhomogeneous media \cite{Merritt18}. In the case of atmospheric turbulence, phase diversity or path-length diversity techniques may be employed to retrieve phase and amplitude information \cite{Fienup82,Fienup13,Shechtman15}. Applications include astronomy, space domain awareness, remote sensing, power beaming, and related areas \cite{Guyon10,Mateen15,Crepp20,Ahn23,Kalensky24}. 

Deciding where to place intensity measurement planes that detect and record diffraction and interference patterns has important implications for practical implementation \cite{Letchev23}. A variety of multi-image phase diversity methods have been developed for near-focal-plane and/or near-pupil-plane measurements using various reconstruction algorithms\cite{Gonsalves82,Teague83,Roddier88,RoddierCubed88,Dean03,Baudat20}. The family of curvature sensors, including both linear and nonlinear signals, often rely on multi-plane, passive intensity measurements with no moving parts \cite{Guyon10}. The principal optical design challenge typically involves creating an optical relay that generates a small diameter beam, long back focal distance, and small opto-mechanical footprint. A small beam diameter is needed to induce diffraction and record multiple images onto a single detector; a long back focal distance is needed to fit beam-splitting components into the instrument; and a small opto-mechanical footprint is desired for practical implementation. These requirements often conflict with one another. 



As a consequence, recording intensity measurements in semi-collimated space to introduce path-length diversity for coherent sensors creates unavoidably large opto-mechanical footprints (tens of centimeters on a side). In practice, the optical bench real-estate needed for pure free-space propagation measurements may not exist. Otherwise, multiple high-speed cameras may be required\cite{Velasco15}, further increasing size, weight, power, complexity and cost. Operating in a convergent beam would address these practical concerns while also {concentrating light into a fewer number of pixels to help combat detector read noise.}

\begin{figure}
    \centering
    \includegraphics[width=0.46\textwidth]{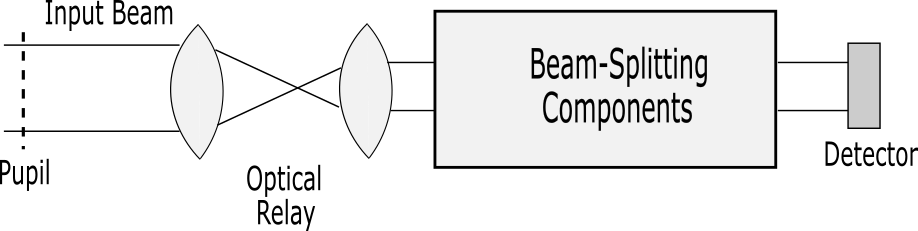} \\
    \includegraphics[width=0.46\textwidth]{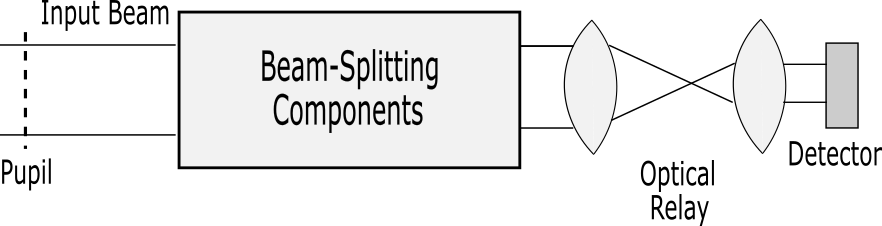} \\
    \includegraphics[width=0.47\textwidth]{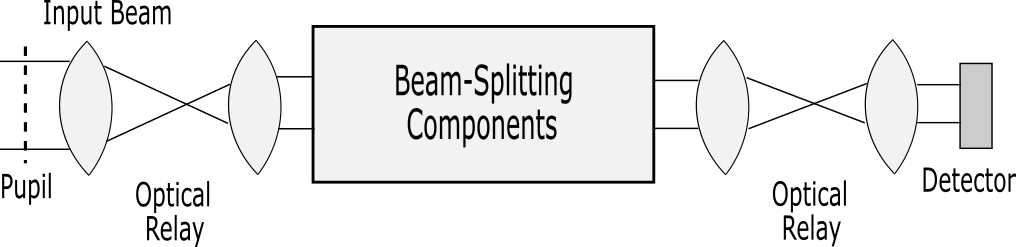} \\
    \includegraphics[width=0.30\textwidth]{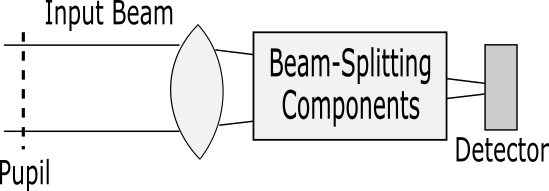}
    \caption{Cartoon level optical design concepts for capturing the sensing channels of a multi-plane wavefront sensor. Beam-splitting components introduce optical path-length differences between channels (individual derivative beams are not shown). Operating in collimated space requires a large opto-mechanical foot-print (top three designs). Operating in a converging beam improves compactness (bottom design).}
    \label{fig:layout}
\end{figure}

As an example, it is instructive to compare the spatial frequency response of the conventional curvature sensor to the non-linear curvature sensor. As a convention, we use $\nu=\sqrt{\nu_x^2 + \nu_y^2}$ for spatial frequencies (cycles / m) so that in two-dimensions
\begin{equation}
    \hat \phi(\nu_x,\nu_y) = \int \int \phi(x,y) e^{- i 2 \pi (\nu_x x + \nu_y y)} dx dy.
\end{equation}
The (linear) curvature sensor estimates the Laplacian of the phase, $\phi(x,y)$, using intensity differences recorded between two planes located near the pupil,
\begin{equation}
    \Delta I(x,y) = C \nabla^2 \phi(x,y),
\end{equation}
where $C$ is a proportionality constant that depends on the optical set-up. The power spectrum of curvature phase estimates is given by
\begin{equation}
    \Phi_{\rm Curvature}(\nu_x,\nu_y) = \; <|\mathrm{FT}(\nabla^2 \phi(x,y))|^2>, 
\end{equation}
where $<>$ is an expectation operator for a statistically isotropic phase field on the pupil. Since $|\mathrm{FT}(\nabla^2 \phi(x,y))| = 4 \pi^2 (\nu_x^2 + \nu_y^2) \hat \phi(\nu_x,\nu_y) = 4 \pi^2 \nu^2 \hat \phi(\nu)$, and $\Phi_{\rm atm}(\nu) = <|\hat \phi(\nu)|^2> \sim \nu^{-11/3}$ is the power spectrum of phase aberrations caused by atmospheric turbulence, the sensitivity of the curvature sensor scales with spatial frequency according to
\begin{equation}
    \Phi_{\rm Curvature}(\nu) \propto \nu^4 \nu^{-11/3} \propto \nu^{1/3}
\end{equation}
assuming Kolmogorov statistics. Differential intensity measurements recorded near the pupil thus emphasize high spatial frequencies \cite{Roddier88}. 

While curvature sensing enhances small-scale turbulence features, operating further from the pupil allows diffraction to smooth high spatial frequency content. Including measurement planes at greater distances (lower Fresnel number), such as with the non-linear curvature sensor, has been shown to significantly improve the sensitivity of curvature sensing \cite{Guyon10,Crepp20,Letchev23}. In the limit of operating in the far field, atmospheric phase distortions begin to imprint their statistics onto the intensity pattern,
\begin{equation}
    \Phi_{\rm Fraunhofer}(\nu) \approx \Phi_{\rm atm}(\nu) \propto \nu^{-11/3},  
\end{equation}
emphasizing lower spatial frequency content which contain more power. As in the case of a four-plane non-linear curvature sensor, the combination of inner-planes and outer-planes leads to a broader, more balanced spatial frequency response. In practice however, the far field generally remains inaccessible using free-space propagation due to unrealistically large $z$-distances. 

Operating a {non-linear curvature WFS closer to focus, e.g. lower Fresnel number} than what a free-space propagation system nominally provides access to, would permit measurements at larger effective propagation distances while also improving signal-to-noise ratio in the presence of read noise. Concentrating the light beam can probe smaller spatial scales, such as in microscopy \cite{Johnson24}, and may reduce computational latency by illuminating a smaller number of pixels. While lab testbeds and field experiments warrant the use of a focusing element to conveniently access larger effective propagation distances, the surface figure of focusing elements creates large phase variations (hundreds of waves of curvature) that preclude common wavefront reconstruction algorithms from converging due to spatial sampling limitations. It is therefore desirable to develop a technique whereby a lens or mirror is used in hardware while not requiring explicit modeling of the optical surface in software.

Curvature sensing and phase diversity near focus have a long history, starting with Roddier 1988\cite{Roddier88} and later nonlinear curvature sensing.\cite{Guyon10} Several nlCWFS implementations have been demonstrated in the lab or using ground-based telescopes, typically in an effort to compare performance with conventional curvature sensing and Shack-Hartmann devices. Mateen et al. built a nlCWFS that used dichroics with fold mirrors to record four intensity measurement planes contemporaneously\cite{Mateen10,Mateen15}. Crass et al. 2012 and Velasco et al. 2015 developed the AOLI instrument, which uses dichroics with a pyramid mirror and two high-speed CCD cameras, to explore combining lucky imaging with AO at the William Herschel Telescope.\cite{Crass12,Velasco15} Crepp et al. 2020 and Letchev et al. 2024 conducted lab experiments using a nlCWFS that split the incoming aberrated beam into four measurement planes based on polarization state.\cite{Crepp20,Letchev24}. Recently, Ahn et al. 2023 installed a nlCWFS at Subaru Observatory that uses dichroics and fold mirrors with the ability to adjust $z$-distance.\cite{Ahn23} While these implementations managed to demonstrate various aspects of the nlCWFS --- including the use of a single detector for multiple out-of-pupil-plane images along with an optical relay to control magnification --- in such cases the designs relied on free-space propagation rather than operating in a convergent beam, resulting in large opto-mechanical footprints.

A nlCWFS that operates in the Fresnel regime using multiple measurement planes within a convergent (or divergent) beam would offer design flexibility and be useful for applications where the AO system real-time controller can keep up with rapid speckle movement to minimize spatial smearing. As we emphasize in this paper, operating in a convergent beam improves compactness of the optical lay-out (Fig.~\ref{fig:layout}). The novelty here is providing a practical framework for multi-plane, nonlinear phase retrieval by creating a simple mapping usable by standard GS (or comparable) implementations with hardware. In $\S$\ref{sec:methods}, we describe retrieval methods used to test a phase diversity sensor that records intensity measurements between the pupil plane and focus. In $\S$\ref{sec:results}, we present simulations {and experimental results} that demonstrate the technique by reconstructing wavefront phase aberrations. Finally, $\S$\ref{sec:summary} provides a summary and concluding remarks. 

\section{Methods}\label{sec:methods}

We aim to develop a lens-based intermediate-focal reconstruction program that uses the Gerchberg-Saxton (GS) or comparable algorithm, modified to incorporate multiple measurement planes. In Letchev et al. 2023\cite{Letchev23}, a comprehensive set of numerical simulations were performed using free-space optical propagation for different non-linear curvature sensor configurations (exploring different numbers of measurement planes and their locations); this work did not include a focusing element. In this study, we develop a practical method for multi-plane wavefront sensing that uses a lens to compress the beam without having to model the lens shape explicitly. The resulting approach avoids algorithm convergence challenges with large phase variations and provides access to greater effective propagation distances while also improving optical system compactness. 

\subsection{Scalar Wave Optics Simulations}\label{sec:model} 

Modeling diffractive wavefront sensors relies on accurate physical optics propagation programs that capture the effects of phase to amplitude conversion. We use scalar wave optics simulations that satisfy the Helmholtz spatial wave equation to model light propagation and sensing. The techniques follow closely that of Letchev et al. 2023\cite{Letchev23} and Potier et al. 2023\cite{Potier23}, which we briefly summarize below, with the exception that a focusing element is introduced into the sensing channel. 

Aberrations are introduced using a single phase screen that follows a Kolmogorov spectrum representative of atmospheric turbulence. The physics of near-field free space propagation is captured using the Fresnel transform-based angular spectrum method \cite{Schmidt10}. We consider a telescope of diameter $D_{\rm tel}=0.5$m experiencing turbulence with a Fried parameter $r_0=15$ cm using monochromatic light of wavelength $\lambda = 633$nm. It is assumed that light from the telescope is relayed down to a $d=2$mm collimated beam. Sensing measurements along the optical axis are performed by a nonlinear curvature sensor that uses four planes located at different $z$-distances for recording local intensity variations \cite{Guyon10}. Phase retrieval is performed using the GS method between measurement planes to reconstruct the wavefront phase \cite{Gerchberg72}. Typically non-linear curvature sensor measurements are taken using pure free-space propagation with flat fold mirrors to steer the derivative beams that provide path-length diversity \cite{Mateen15,Ahn23}. In this study, we assume that a lens is introduced to focus the beam prior to beam splitting (Fig.~\ref{fig:layout}). 

\begin{figure}
    \centering
    \includegraphics[trim=0.7in 0.2in 0.2in 0.2in, clip, width=0.5\linewidth]{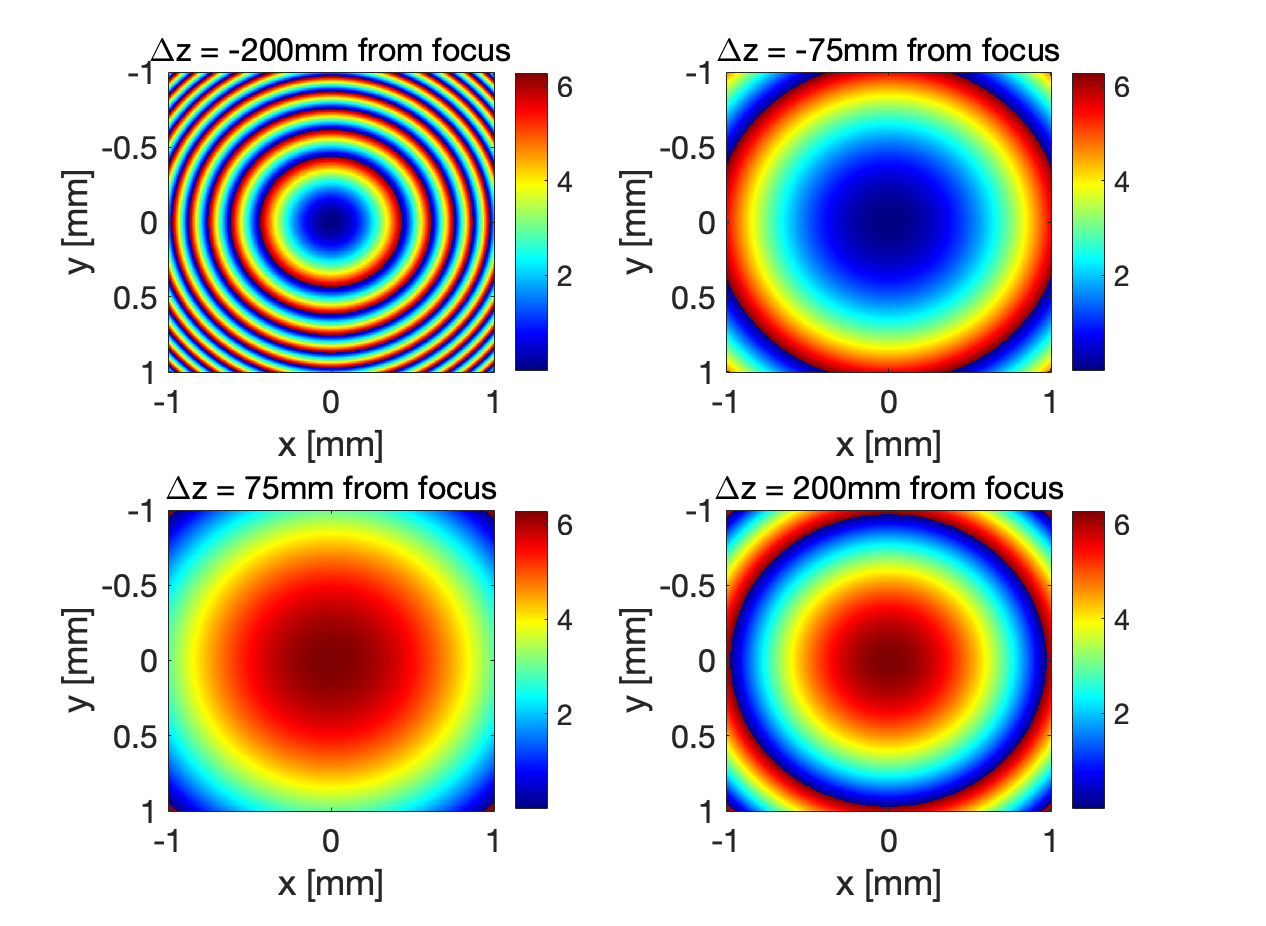}
    \caption{Quadratic kernel phase, $\Phi_{\rm kernel}$ (radians), appearing in the Collins/Fresnel propagation integral for a thin lens plus free-space gap plotted across the entrance pupil for several axial distances. The model assumes a wavelength of $\lambda=633$ nm, beam diameter of $d=2$ mm, and $f=300$ mm focal length.} 
    \label{fig:phase_lens}
\end{figure}

Initial attempts at modeling the refractive surface of the lens resulted in divergent algorithm behavior due to large phase variations induced by the lens and resulting need for unrealistically high spatial sampling. For a thin lens followed by propagation distance $z$, the Fresnel propagator contains a quadratic phase factor on the input (pupil) coordinates ($x_1, y_1$). We define its argument for paraxial systems as 
\begin{eqnarray}
\Phi_{\rm kernel}(x_1,y_1;z)  &=& - \frac{k}{2f} (x_1^2 + y_1^2) + \frac{k}{2 z} (x_1^2 + y_1^2) \nonumber \\
             &=& \frac{k}{2} \left( \frac{1}{z} - \frac{1}{f}\right)(x_1^2 + y_1^2) = \frac{k}{2} \left( \frac{f - z}{zf}\right)(x_1^2 + y_1^2),  
\end{eqnarray}
where $k$ is wavenumber and $f$ is the lens focal length. (See \S \ref{sec:appendix} for details using the ABCD Collins formalism.) Figure~\ref{fig:phase_lens} plots $\Phi_{\rm kernel}(x_1,y_1)$ across the entrance pupil to illustrate how rapidly the propagation kernel oscillates and why explicit modeling of the lens term can create severe sampling and aliasing demands. These high frequency structures exacerbate the challenge of phase retrieval motivating the need for a work-around solution.

For example, in order for the GS algorithm to converge, the amplitude at each measurement plane is applied as a known boundary condition on the complex field; the phase is then iteratively updated following Fresnel propagation between planes. However, when a lens is introduced into the optical train, the phase varies strongly in a convergent beam requiring an unrealistic number of pixel samplings for the algorithm to avoid aliasing, which in turn introduces memory and speed limitations. In practice, we find that modeling the lens surface limits the ability of wavefront reconstruction algorithms to converge. Instead, it may be desirable to by-pass the need to model powered optical surfaces entirely. We develop below a technique based on paraxial diffraction theory that simplifies the problem of phase retrieval using powered optical elements.  

\subsection{Beam Compression and Magnification}\label{sec:compression}

Our recommended procedure for working in a focused beam involves correcting for image demagnification in software. Taking into consideration the competing effects of diffraction versus convergent beams, we aim to perform wavefront reconstructions as if the lens were not included in the optical train, by ``remagnifying'' the measured (focused) intensity patterns using numerical methods. 

To incorporate the effects of a lens (or mirror) of focal length $f$, which enables proximal access to the far field, we define an effective travel distance,
\begin{equation}\label{eqn:zeff}
    z_{\rm eff} \equiv \frac{z f}{f - z}
\end{equation}
where $z=0$ references the location of the lens, which we assume is co-located at the pupil. This equation, which is valid for $z < f$, is designed such that $z_{\rm eff} \rightarrow 0$ when $z \rightarrow 0$ and $z_{\rm eff} \rightarrow \infty$ when $z \rightarrow f$. Effectively, $z_{\rm eff}$ is the distance that light would have traveled in free space to accumulate the same amount of diffraction (curvature, phase spread, expansion) as it does at a distance $z$ after passing through the lens.\footnote{Gaussian beams would expand based on wavelength and the beam waist.} The Appendix (\S \ref{sec:appendix}) provides a derivation and justification for Equation~\ref{eqn:zeff}, which is valid for paraxial optical systems.

In the absence of a lens, the angular size of a non-focused, free-space beam expands with $z$-distance as $\theta \sim \lambda / d$, and the linear diameter of the beam scales as 
\begin{equation}
    B_{\rm fs}(z) \propto \frac{\lambda}{d} z
\end{equation}
to within a constant or order unity. This occurs in the reference frame of the reimaged pupil of the telescope, which has a collimated beam diameter $d$ instead of $D_{\rm tel}$. Incorporating the effects of a lens, the linear diameter of the beam expands with $z$-distance according to
\begin{equation}
    B_{\rm lens}(z) \propto \frac{\lambda}{d} z_{\rm eff}(z).
\end{equation}
We calculate a unitless scaling factor, $S$, that magnifies focused images based on the location of the measurement planes and lens focal length (see $\S$\ref{sec:appendix} for details), 
\begin{equation}\label{eqn:Sfac}
    S \equiv \frac{B_{\rm lens}(z)}{B_{\rm fs}(z)} = \frac{z_{\rm eff}}{z} = \frac{f}{f-z} = \frac{1}{1 - z/f}. 
\end{equation}
In the limit $z \rightarrow 0$, $S=1$ indicating no magnification at the pupil. In the limit $z \rightarrow f$, $S \rightarrow \infty$ and the magnification formally diverges to infinity: by design this feature offers access to the far field. In the limit $f \rightarrow \infty$, the lens/mirror has negligible focusing power and $S = 1$ indicating no magnification relative to free-space propagation. For small $z << f$, $S \approx 1 + z/f$ describing near-field linear behavior. 

\begin{figure}
    \centering
    \includegraphics[width=0.48\textwidth]{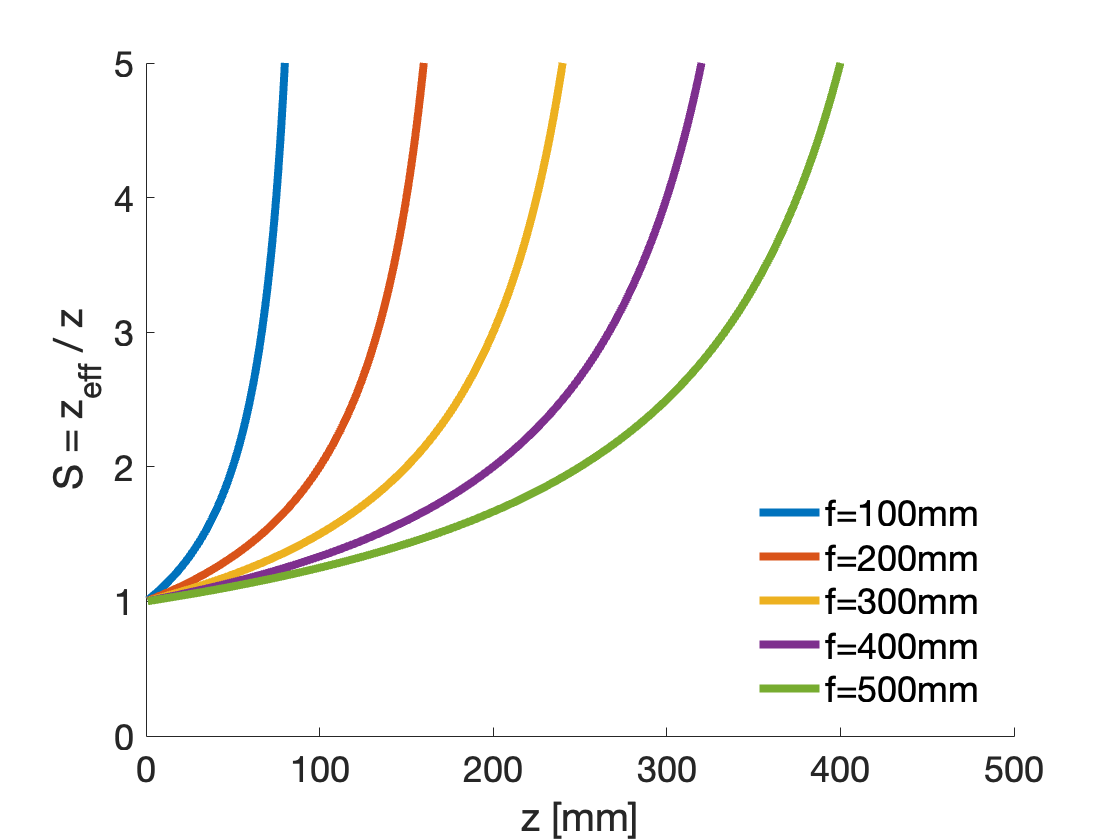}
    \caption{Image scaling factor, $S = z_{\rm eff} / z$, evaluated for several different focal lengths. A focusing optic provides access to many multiples of the linear distance, $z$, used for creating optical path-length differences.} 
    \label{fig:scale_factor}
\end{figure}

Since $S = z_{\rm eff} / z > 1$ and $z_{\rm eff} > z$, the use of a focusing element provides convenient access to the Fresnel and Fraunhofer regimes by generating $z_{\rm eff}$ values that are many multiples of $z$ (Fig.~\ref{fig:scale_factor}). While the scale factor increases rapidly as $z \rightarrow f$, large $S$ values can also result in unmanageably large images (or plate-scales). In practice, it is necessary to work with modest $S$ values to preserve spatial sampling. Near focus, individual measurement planes should be separated by at minimum the system depth of focus to offer sufficient diversity. Far field images at focus ($z=f$) may also be incorporated into the phase retrieval process with appropriate changes to the reconstruction algorithm, namely using Fourier transforms instead of Fresnel transforms relative to the pupil when $S$ becomes undefined. The technique can work over a broad range of collimated beam diameters, provided that the system f-number offers adequate image quality and spatial sampling requirements are met. 

In practice, it may be convenient to work with images recorded on the other side of focus due to mechanical space constraints. Operating a multi-plane sensor with divergent beams is also governed by Eqn.~\ref{eqn:zeff} and Eqn.~\ref{eqn:Sfac} where $z$ is measured from the lens. Negative $z_{\rm eff}$ values are just book-keeping to indicate that planes are located beyond focus as the light propagates from left to right. The scaling factor, $S$, also becomes negative indicating an inverted image as expected; wavefront reconstruction programs would need to take this into account either in software or when establishing deformable mirror interaction relationships.

\subsection{Phase Retrieval}\label{sec:retrieval}

Recorded intensities for each $z$-distance are stretched by the $S$ value and interpolated such that the reconstruction algorithm interprets each image as a free-space propagation. For example, in Matlab we use the \texttt{imresize( )} function to modify the scale of images (Fig.~\ref{fig:theory_images}). Images may be centered prior to stretching to exclude tip/tilt; alternatively, the methods described in Abbott et al. 2025\cite{Abbott25} may be employed to retrieve tip/tilt modes. The GS or comparable method is then applied to reconstruct the wavefront and finally the phase is unwrapped at the pupil (Huerta et al. 2025) \cite{Huerta25}.

\subsection{Normalized Coordinates Interpretation}\label{sec:normalized}
We briefly discuss an equivalent normalized coordinate interpretation based on Fresnel number and consider the resulting sampling requirements. A thin lens followed by a free-space gap is a first-order (paraxial) optical system that can be described by an ABCD matrix \cite{Shakir15}. Using the Collins diffraction integral ($\S$\ref{sec:appendix}), the field at an axial distance $z$ downstream of a thin lens of focal length $f$ can be written as a Fresnel-type integral with a quadratic phase kernel.\cite{Collins70} For $z < f$, this lens-plus-gap system is equivalent (up to a deterministic output phase factor and an amplitude scaling) to free-space propagation over an effective distance $z_{\rm eff}$, evaluated at a scaled transverse coordinate $S \equiv \frac{z_{\rm eff}}{z}=\frac{f}{f-z}$. This mapping is derived in the Appendix ($\S$\ref{sec:appendix}) and yields the intensity relation
\begin{equation}
    I_{\rm lens}(x,y; z) = S^2 I_{\rm fs}(Sx, Sy;z_{\rm eff}),
\end{equation}
so that the measured intensity patterns in a convergent beam are geometrically demagnified versions of the corresponding free-space patterns evaluated at 
$z_{\rm eff}$.

\subsection{Fresnel Number Equivalence}
To quantify this equivalence in the usual paraxial diffraction parameterization, we define the Fresnel number for a pupil of radius $a=d/2$ and propagation distance $Z$ as
\begin{equation}
    N_F(Z) \equiv \frac{a^2}{\lambda Z}.
\end{equation}
Under the mapping above, a lens-plus-gap measurement at physical distance $z$ corresponds to a free-space distance $Z=z_{\rm eff}$, and hence the effective Fresnel number is
\begin{equation}
    N_F^{\rm eff} = N_F(z_{\rm eff}) = \frac{a^2}{\lambda z_{\rm eff}} = \frac{a^2}{\lambda}\frac{(f-z)}{fz}.
\end{equation}
Thus the lens does not alter the underlying paraxial diffraction physics; instead, it provides access to small $N_F^{\rm eff}$ (large effective propagations) using a compact physical path length $z$.

\subsection{Sampling and Aliasing Considerations}\label{sec:sampling}
While the continuous theory is equivalent, discrete implementations of phase retrieval require adequate sampling of the rapidly varying quadratic phase factors that appear in Fresnel propagation. A conservative condition to avoid aliasing is that the maximum phase increment between adjacent pupil samples remain below $\pi$. For a quadratic phase of the form
\begin{equation}
    \phi(x) = \frac{k}{2 Z} x^2,
\end{equation}
the local phase slope is $\partial \phi / \partial x = kx/Z$, so at the pupil edge, $x=a$, a no-aliasing criterion is 
\begin{equation}
    \Delta x \left \| \frac{\partial \phi}{\partial x} \right \|_{x=a} = \frac{k a}{Z} \Delta x \lesssim \pi 
\end{equation}
or
\begin{equation}
    \Delta x \lesssim \frac{\lambda Z}{2 a}
\end{equation}
Setting $Z = z_{\rm eff}$ yields the sampling requirement in the normalized-coordinate (free-space-equivalent) representation:
\begin{equation}
    \Delta x \lesssim \frac{\lambda z_{\rm eff}}{2 a}. 
\end{equation}
This shows explicitly that the sampling burden is governed by the same effective propagation distance (equivalently the same effective Fresnel number) regardless of whether one models the lens-plus-gap system directly or uses the free-space-equivalent mapping ($\S$\ref{sec:appendix}).

\subsection{Methods Comparison}\label{sec:comparison}
In summary, there are (at least) two comparable ways to incorporate the lens-plus-gap mapping into a multi-plane phase retrieval pipeline:
\begin{enumerate}
    \item Matrix propagation operators (non-FFT-based, Shakir et al. 2015): Implement canonical transforms using linear algebra methods to propagate complex fields between planes with flexible sampling.\cite{Shakir15}
    \item Rescaled intensity constraints (FFT-based, this work): Resample measured intensities to synthesize equivalent free-space diffraction patterns onto a common computational grid, then apply FFT-based propagation.
\end{enumerate}
The usage of matrix methods versus intensity scaling are equivalent in continuous paraxial theory (see $\S$\ref{sec:appendix}). Selection of a method thus depends on the application, requiring a balance between noise properties, accuracy, and computational latency. In wavefront sensing, operating in strong turbulence or at short wavelengths (small $r_0$) drives spatial sampling demands, creating a complex trade-space and design landscape. 

For instance, while FFT-based Fresnel propagation benefits from the simplicity of a fixed-grid and inverse-transform consistency, it suffers from periodic boundary effects that may cause aliasing when left unchecked. The aforementioned Shakir et al. 2015 efficient matrix algorithm offers flexible input/output sampling while eliminating wrap-around artifacts caused by FFT-methods, so is inherently anti-aliasing. Intensity interpolation may also alter pixel-level noise statistics (e.g., inducing correlations), though this can be largely mitigated by ensuring adequate detector sampling and appropriate pixel-weighting in the projection step when operating at low illumination levels. Considering these trade-offs, intensity scaling may offer a pragmatic implementation that provides a low-friction bridge from convergent-beam measurements to standard free-space multi-plane reconstructors, which is valuable for AO systems and may be attractive for rapid prototyping. Our aim is not to argue that one propagator family is universally better, but to provide a practical approach that allows fixed-grid GS-like pipelines to work with convergent-beam data without implementing a new wave-optics operator.


\begin{figure*}
    \centering
    \includegraphics[trim=0.0in 0.0in 0.0in 0.30in, clip, width=0.98\linewidth]{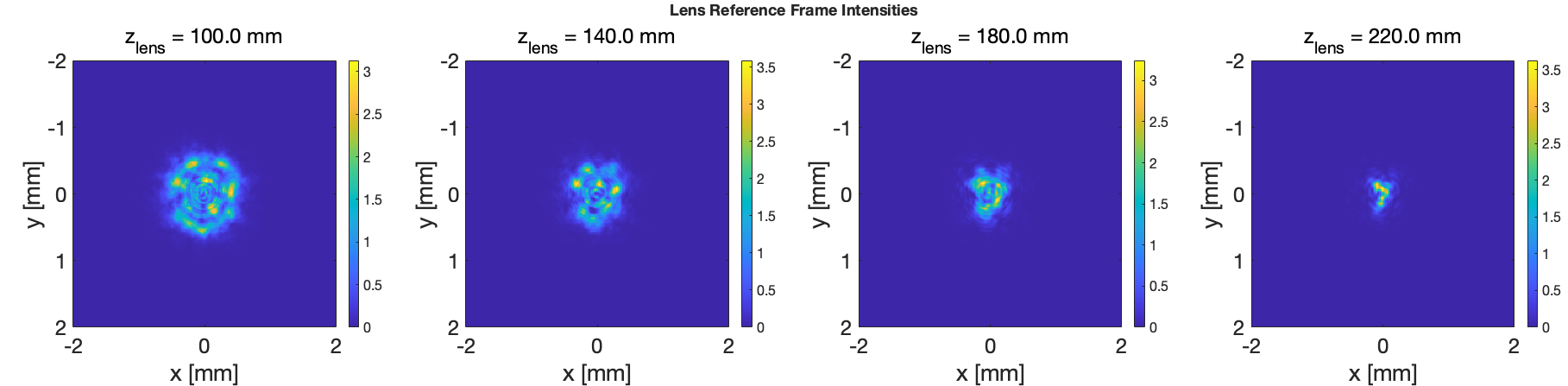}
    \includegraphics[trim=0.0in 0.0in 0.0in 0.30in, clip, width=0.98\linewidth]{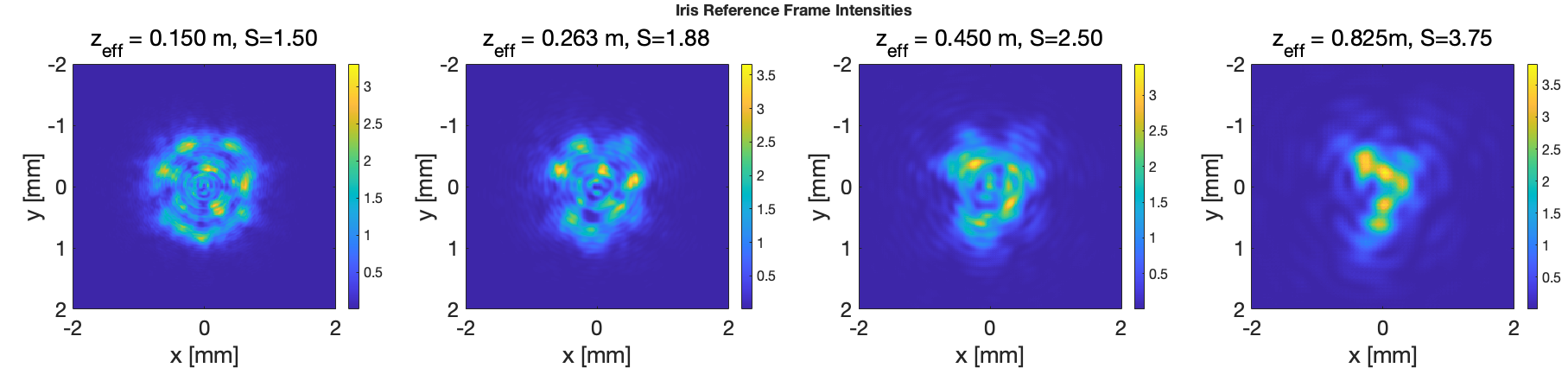}
    \caption{Simulated $f=300$ mm lens images (top row) and rescaled images (bottom row) including atmospheric turbulence. Measured $z$-distances are listed in the top panel while $z_{\rm eff}$ and $S=z_{\rm eff}/z$ are listed in the bottom panel. Color-bars are in units of counts.}
    \label{fig:theory_images}
\end{figure*}

\begin{figure*}
    \centering
    \includegraphics[trim=0.0in 0.0in 0.0in 0.3in, clip, width=0.98\linewidth]{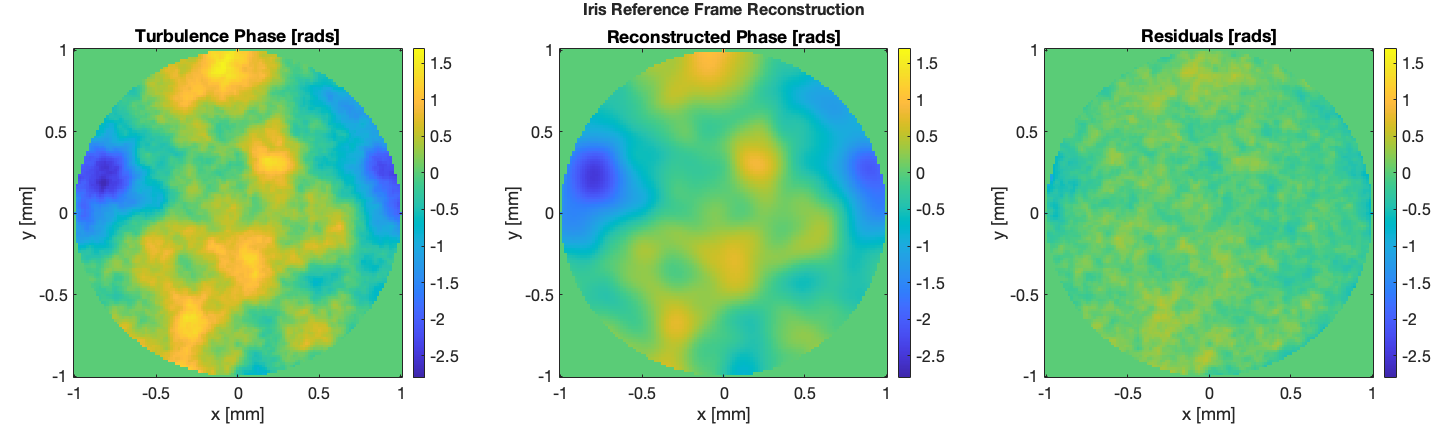}
    \caption{Wavefront reconstruction of a Kolmogorov phase screen representing atmospheric turbulence using simulations.}
    \label{fig:reconstruction}
\end{figure*}

\section{Results}\label{sec:results}

As a demonstration of the process, we perform a wavefront reconstruction in simulations using the parameters listed in $\S$\ref{sec:methods} with an $f=300$ mm lens. Distances along the optical axis, $z = [100, 140, 180, 220]$ mm corresponding to $z_{\rm eff} = [150, 263, 450, 825]$ mm, were chosen to offer a balance of sensitivity to different spatial frequencies and manageable $S$ values (Fig.~\ref{fig:theory_images}). After image scaling, spatial sampling was interpolated to create a 3.45 $\mu$m pixel pitch, which matches a camera in the lab that we used for experimental validation of the technique. Tip/tilt was not removed in software but was reconstructed as part of the electric field. Figure~\ref{fig:reconstruction} shows results using the methods described above in the reference frame of the $d=2$ mm beam relay conjugate to the telescope. We find an excellent match between the injected wavefront and reconstructed phase. Residuals are comprised primarily of high spatial frequencies producing an RMS WFE of 0.03 waves (0.17 radians). 

\begin{figure*}
    \centering
    \includegraphics[trim=0.0in 0.0in 0.0in 0.7in, clip, width=0.98\linewidth]{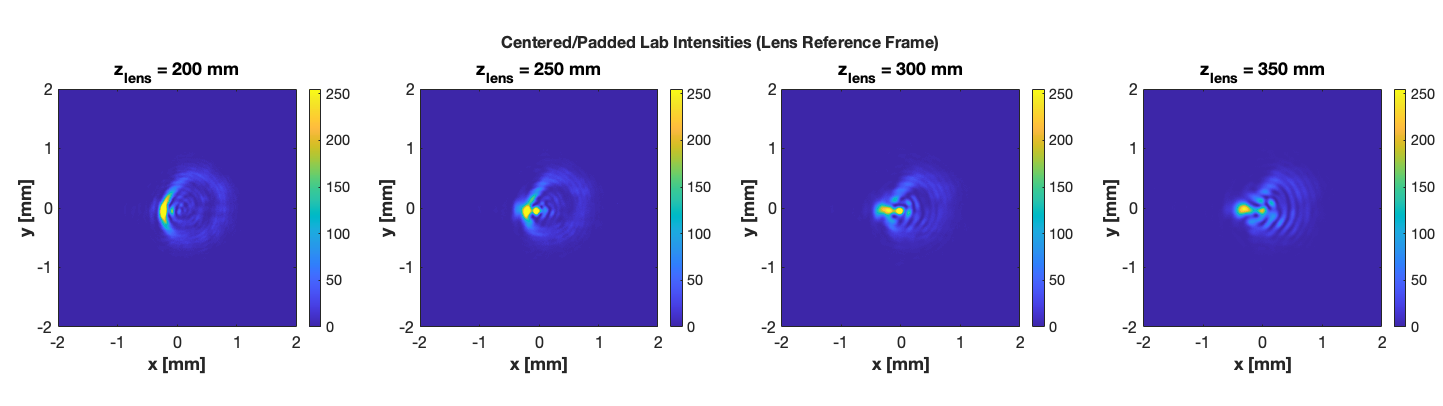}
    \includegraphics[trim=0.0in 0.0in 0.0in 0.7in, clip, width=0.98\linewidth]{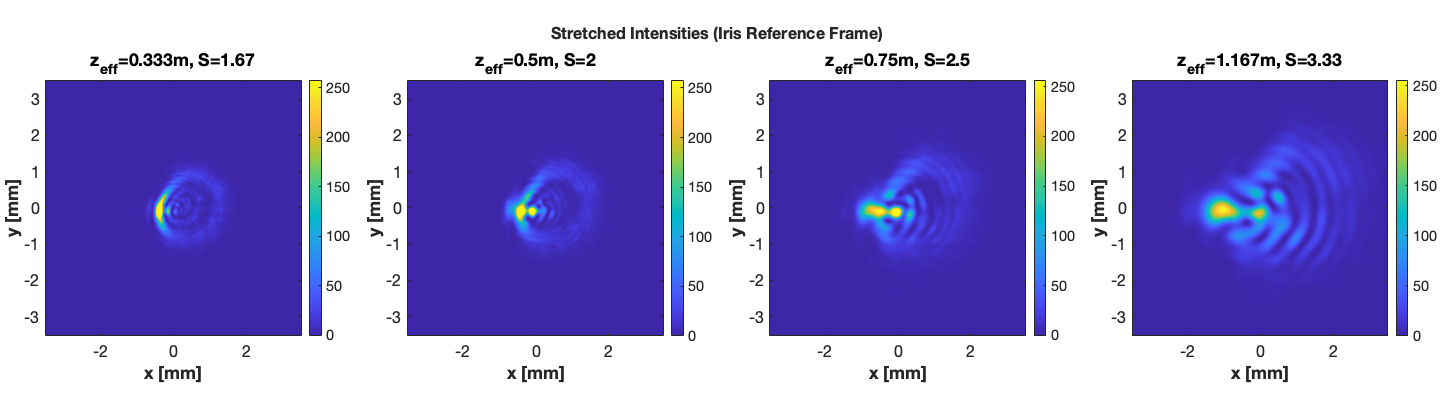}
    \caption{Laboratory $f=500$ mm lens images (top row) and rescaled images (bottom row) of pure coma. Measured $z$-distances are listed in the top panel while $z_{\rm eff}$ and $S=z_{\rm eff}/z$ are listed in the bottom panel. Color-bars are in units of counts.}
    \label{fig:lab_images}
\end{figure*}

\begin{figure*}
    \centering
    \includegraphics[trim=0.0in 0.36in 0.2in 0.3in, clip, width=0.43\linewidth]{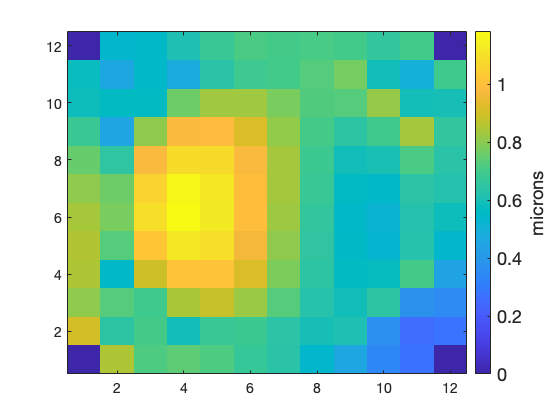}
    \includegraphics[trim=0.3in 0.34in 0.0in 0.3in, clip, width=0.43\linewidth]{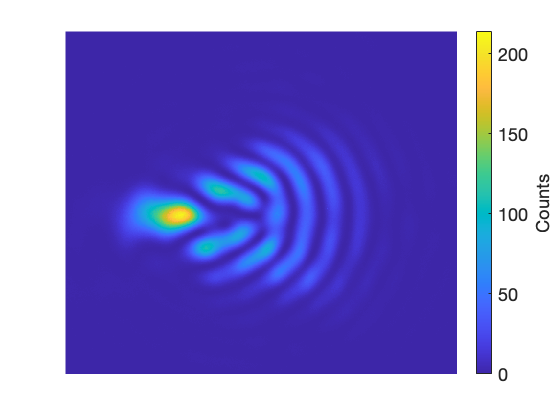} \\
    \includegraphics[trim=0.3in 0.4in 0.1in 0.3in, clip, width=0.42\linewidth]{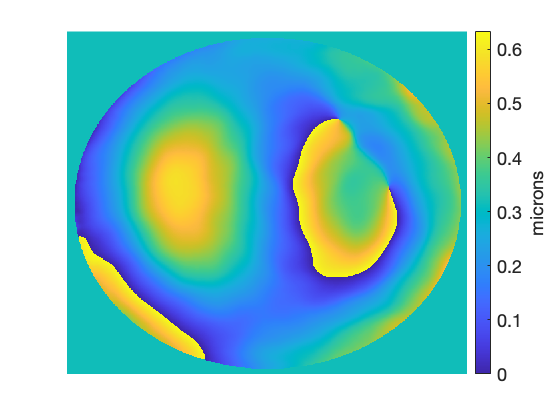} 
    \includegraphics[trim=0.3in 0.4in 0.1in 0.3in, clip, width=0.42\linewidth]{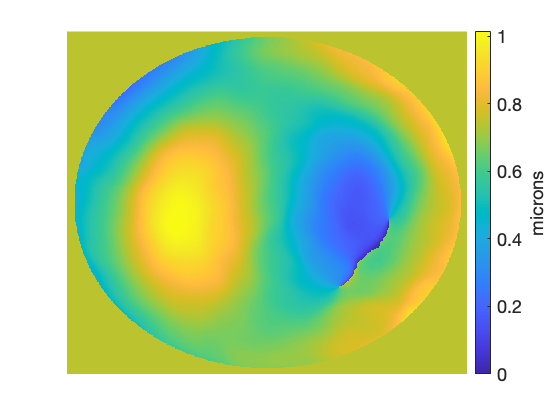} 
    \caption{First wavefront reconstruction in the lab using a lens to compress an aberrated beam ($\lambda = 633$ nm). Deformable mirror surface commands of horizontal coma using $12 \times 12$ actuators to induce a known aberration (top-left). Focal plane image at $z=f$ showing coma (top-right). Reconstructed phase (bottom-left). Unwrapped reconstructed phase (bottom-right).}
    \label{fig:coma}
\end{figure*}

To further substantiate the technique, we introduced an $f=500$ mm lens into an active sensing experiment in the AO beam control testbed at Notre Dame (Fig.~\ref{fig:lab_images}). The testbed is similar to is similar to Crepp et al. 2020\cite{Crepp20} and includes turbulence generation and multiple wavefront sensing channels for comparing different sensing technologies. Included in the beam is a continuous face-sheet Boston Micromachines $12 \times 12$ deformable mirror that we used to introduce known aberration shapes for calibration and testing. In the experiment, nearly two waves (peak-to-valley) of horizontal coma was created by the deformable mirror using a HeNe laser ($\lambda=633$ nm). Images were recorded at different distances, $z = [200, 250, 300, 350]$ mm, to create phase diversity. Figure~\ref{fig:coma} shows a laboratory demonstration of wavefront reconstruction before and after phase unwrapping, which was performed using the ``Fast2D'' technique by Herraez et al. 2002\cite{Herraez02}. The unwrapped reconstruction clearly and unambiguously recovers the coma aberration demonstrating the validity of the technique when using a lens in practice. Importantly, we did not provide any information to the reconstruction algorithm about the lens shape other than the specified focal length (i.e not glass type, index of refraction, radii of curvature, or thickness); this form of regularization should be robust for paraxial systems ($\S$\ref{sec:appendix}). 

\begin{figure}
    \centering
    \includegraphics[width=0.25\linewidth]{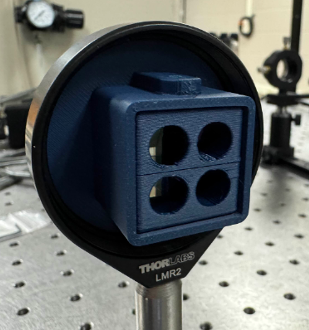}
    \caption{Miniaturized multi-plane sensor housed in a standard 2-inch diameter Thorlabs lens mount. Shown in the image is the four-channel output. A lens is used upstream to create a convergent beam. Internal beam splitters create path-length diversity.}
    \label{fig:swap}
\end{figure}

Figure~\ref{fig:swap} shows an image from the lab of a small size and small weight nonlinear curvature sensor. The device, which receives a monochromatic convergent beam, fits within a standard 2-inch diameter Thorlabs lens mount and may be used with short focal length beams. Internal to the sensor are a series of beam-splitters that separate the light based on amplitude. This perspective shows the exit window that outputs four derivative beams. Other configurations and beam-splitting techniques are possible depending on application and detector geometry. The device was used to develop a data analysis pipeline that verifies the methods described in $\S\ref{sec:methods}$. 

\section{Summary}\label{sec:summary}

We have developed an optical engineering technique that significantly reduces the size, weight, complexity and cost required for multi-plane diffractive wavefront sensors like the nonlinear curvature sensor. The approach incorporates a focusing element in hardware, yet does not require explicit modeling of the optical component surface in software. With only a modest increase in reconstructor complexity, this design feature greatly simplifies experiments while bypassing the need to perform phase retrieval amid hundreds of waves of phase variation. As a result, building compact optical sensors that use diffraction and path-length diversity for telescope systems becomes more feasible. 


\section{Disclosures}\label{sec:disclosure}
 
The authors declare that they have no financial interests, commercial affiliations, or other competing interests that could have influenced the work reported in this paper. 

\section{Data Availability}\label{sec:data}
 
Data used to create the figures for this manuscript are available upon request from the authors. 

\section{Appendix}\label{sec:appendix}

We show that a lens plus propagation gap with image scaling is equivalent to free-space propagation using the ABCD Collins diffraction formalism  \cite{Collins70}. For compactness, we work in one-dimension ($x$); the two-dimensional result follows by separability $(x,y)$ with each axis. For a first-order (paraxial) system with ray-transfer matrix
\[
M = 
\begin{pmatrix}
A & B\\[2pt]
C & D
\end{pmatrix},
\]
the Collins diffraction integral gives the output field $U_2(x_2)$ in terms of the input field
$U_1(x_1)$ as
\begin{equation}
U_2(x_2)
=
\frac{e^{ikL}}{i\lambda B}\;
\exp\!\left(i\frac{k D}{2B}x_2^2\right)
\int_{-\infty}^{\infty}
\exp\!\left[i\frac{k A}{2B}x_1^2
-\; i\frac{k}{B}x_1 x_2 \right]
\,U_1(x_1)\,dx_1.
\label{eq:collins_general}
\end{equation}
Here $k = 2\pi/\lambda$ and $L$ is an overall path length (its contribution is a global phase factor, which we suppress below). See Fienup 2024 for details.\cite{Fienup24}.

Consider a thin lens of focal length $f$ at the pupil, followed by a free-space gap $z$:
\[
M_{\rm lens+gap} =
\underbrace{\begin{pmatrix}
1 & z\\[2pt]
0 & 1
\end{pmatrix}}_{\text{free space }z}\;
\underbrace{\begin{pmatrix}
1 & 0\\[2pt]
-\tfrac{1}{f} & 1
\end{pmatrix}}_{\text{thin lens }f}
=
\begin{pmatrix}
1 - \tfrac{z}{f} & z\\[2pt]
-\tfrac{1}{f} & 1
\end{pmatrix}.
\]
Thus
\[
A = 1-\frac{z}{f},\qquad
B = z,\qquad
C = -\frac{1}{f},\qquad
D = 1.
\]
Inserting these into~\eqref{eq:collins_general} gives the field at the detector plane,
\begin{equation}
U_{\rm lens}(x_2)
=
\frac{1}{i\lambda z}\;
\exp\!\left(i\frac{k}{2z}x_2^2\right)
\int_{-\infty}^{\infty}
\exp\!\left[i\frac{k}{2z}\Bigl(A x_1^2 - 2 x_1 x_2\Bigr)\right]
\,U_1(x_1)\,dx_1,
\label{eq:collins_lens_gap}
\end{equation}
with $A = 1 - z/f$. For comparison, pure free-space propagation over a distance $Z$ has ABCD matrix
$\bigl[\begin{smallmatrix}1 & Z\\ 0 & 1\end{smallmatrix}\bigr]$, so $A=D=1$, $B=Z$, $C=0$, and \eqref{eq:collins_general} reduces to
\begin{equation}
U_{\mathrm{fs}}(x_2;Z)
=
\frac{1}{i\lambda Z}\;
\exp\!\left(i\frac{k}{2Z}x_2^2\right)
\int_{-\infty}^{\infty}
\exp\!\left[i\frac{k}{2Z}\Bigl(x_1^2 - 2 x_1 x_2\Bigr)\right]
\,U_1(x_1)\,dx_1.
\label{eq:collins_free_space}
\end{equation}
We now choose an \emph{effective} free-space distance $z_{\mathrm{eff}}$ and a scaled output coordinate so that the kernel in~\eqref{eq:collins_lens_gap} matches the free-space kernel in~\eqref{eq:collins_free_space}. Define
\[
A = 1-\frac{z}{f},\qquad
z_{\mathrm{eff}} \equiv \frac{B}{A} = \frac{z}{1-\tfrac{z}{f}} = \frac{z f}{f - z},
\]
and introduce a scaled detector coordinate in the convergent beam,
\[
x_2 = A\,x_2',
\qquad\text{so that}\qquad x_2' = \frac{x_2}{A}.
\]
Substituting $x_2 = A x_2'$ into the lens expression \eqref{eq:collins_lens_gap} gives
\begin{equation}
U_{\rm lens}(A x_2')
=
\frac{1}{i\lambda z}\;
\exp\!\left(i\frac{k A^2}{2z}x_2'^2\right)
\int_{-\infty}^{\infty}
\exp\!\left[i\frac{k}{2z}\Bigl(A x_1^2 - 2 A x_1 x_2'\Bigr)\right]
\,U_1(x_1)\,dx_1.
\label{eq:lensgap_scaled_x2prime}
\end{equation}
Using $z_{\mathrm{eff}} = z/A$ we have
\[
\frac{A}{2z} = \frac{1}{2z_{\mathrm{eff}}},\qquad
\frac{A}{z} = \frac{1}{z_{\mathrm{eff}}},
\]
so the exponential inside the integral in \eqref{eq:lensgap_scaled_x2prime} becomes
\[
\exp\!\left[i\frac{k}{2z}\Bigl(A x_1^2 - 2 A x_1 x_2'\Bigr)\right]
=
\exp\!\left[i\frac{k}{2z_{\mathrm{eff}}}\Bigl(x_1^2 - 2 x_1 x_2'\Bigr)\right].
\]
This is precisely the free-space kernel at distance $z_{\mathrm{eff}}$ with output coordinate $x_2'$.

Next we compare the prefactors. Since $z = A z_{\mathrm{eff}}$,
\[
\frac{1}{i\lambda z} = \frac{1}{i\lambda A z_{\mathrm{eff}}} = \frac{1}{A}\,\frac{1}{i\lambda z_{\mathrm{eff}}}.
\]
After splitting the quadratic phase on $x_2'$ as
\[
\exp\!\left(i\frac{k A^2}{2z}x_2'^2\right)
=
\exp\!\left(i\frac{k}{2z_{\mathrm{eff}}}x_2'^2\right)
\exp\!\left(i\frac{k}{2}\Bigl[\frac{A^2}{z} - \frac{1}{z_{\mathrm{eff}}}\Bigr]x_2'^2\right),
\]
and using $z_{\mathrm{eff}}=z/A$, the extra factor is
\[
\Phi_{\mathrm{out}}(x_2') \equiv \frac{k}{2}\,\frac{A^2 - A}{z}\,x_2'^2,
\]
which is a known quadratic phase on the detector coordinate. Putting these pieces together, we rewrite \eqref{eq:lensgap_scaled_x2prime} as
\begin{equation}
U_z(A x_2')
=
\frac{1}{A}\,\exp\!\bigl(i\Phi_{\mathrm{out}}(x_2')\bigr)\,
\left[
\frac{1}{i\lambda z_{\mathrm{eff}}}\;
\exp\!\left(i\frac{k}{2z_{\mathrm{eff}}}x_2'^2\right)
\int_{-\infty}^{\infty}
\exp\!\left[i\frac{k}{2z_{\mathrm{eff}}}\Bigl(x_1^2 - 2 x_1 x_2'\Bigr)\right]
\,U_1(x_1)\,dx_1
\right].
\label{eq:lensgap_vs_freespace}
\end{equation}
The bracketed term is exactly $U_{\mathrm{fs}}(x_2';z_{\mathrm{eff}})$ from \eqref{eq:collins_free_space}. Dropping the global propagation phase $e^{ikL}$ for simplicity, we obtain the compact, side-by-side relation
\begin{equation} \label{eq:equivalence_side_by_side}
U_{\rm lens}(A x_2')
=
\frac{1}{A}\,\exp\!\bigl(i\Phi_{\mathrm{out}}(x_2')\bigr)\,
U_{\mathrm{fs}}(x_2';z_{\mathrm{eff}}),
\end{equation} 
where $z_{\mathrm{eff}} = \frac{z f}{f - z}$, and $A = 1-\frac{z}{f}$. For ease of interpretation, we defined the normalized, unitless scaling factor $S \equiv \frac{z_{\rm eff}}{z} = \frac{1}{A} = (1 - \frac{z}{f})^{-1} = \frac{f}{f-z}$ in the main body of the article. 

Taking the modulus squared of \eqref{eq:equivalence_side_by_side} shows that the intensity at the detector plane in the lens system is related to the free-space intensity at $z_{\mathrm{eff}}$ by
\[
I_{\rm lens}(A x_2'; z) = |U_{\rm lens}(A x_2'; z)|^2
= \frac{1}{A^2}\,\bigl|U_{\mathrm{fs}}(x_2';z_{\mathrm{eff}})\bigr|^2
= \frac{1}{A^2}\,I_{\mathrm{fs}}(x_2';z_{\mathrm{eff}}) = S^2 \,I_{\mathrm{fs}}(x_2';z_{\mathrm{eff}}),
\]
since the phase factor $\exp(i\Phi_{\mathrm{out}})$ has unit modulus. Equivalently, writing $x_2 = A x_2' = \frac{1}{S} x_2'$ or $x_2' = x_2/A = S x_2$,
\begin{equation}
I_{\rm lens}(x_2, z) = \frac{1}{A^2}\;
I_{\mathrm{fs}}\!\Bigl(\frac{1}{A} x_2; z_{\mathrm{eff}}\Bigr) = S^2 I_{fs}\!\Bigl(S x_2; z_{\mathrm{eff}}\Bigr),
\label{eq:intensity_stretch}
\end{equation}
or
\begin{equation}
    I_{\rm fs}(x_2'; z_{\rm eff}) = A^2 I_{\rm lens}(Ax_2'; z) = \frac{1}{S^2}I_{\rm lens}\Bigl(\frac{x_2'}{S}; z\Bigr) = \frac{1}{S^2} I_{\rm lens} \Bigl(x_2;z\Bigr)
\end{equation}
For intensity measurements, the lens plus gap system produces at a distance $z < f$ exactly the same pattern that would be observed after free-space propagation over a longer distance $z_{\mathrm{eff}}$, but magnified laterally. In practice, we start from the smaller measured patterns $I_z$ and numerically stretch them by $S$ to synthesize the larger free-space patterns $I_{\rm fs}$. In other words, as the sensor is moved toward the focal plane, the effective propagation distance $z_{\mathrm{eff}}$ grows without bound, and the recorded near-focal images can be interpreted as geometric versions of the far-field pattern evaluated at $z_{\mathrm{eff}}$ up to an overall calibration factor $1/S^2$. The quadratic phase $\Phi_{\mathrm{out}}$ is deterministic and does not affect intensity statistics; it can be ignored when only $|U|^2$ is used. In two transverse dimensions $(x,y)$, the kernel in \eqref{eq:collins_lens_gap} is separable, and the same relations hold with $x\to(x,y)$. The effective distance $z_{\mathrm{eff}}$ and stretch factor $S = \frac{1}{A} = f/(f-z)$ are common to both axes.

\section{Acknowledgments}\label{sec:acknowledge}
 

We thank D. Angelica Huerta for unwrapping the phase shown in Figure~\ref{fig:coma}. This work was supported in part by Air Force Office of Scientific Research (AFOSR) grant number FA9550-22-1-0435 and the Joint Directed Energy Transition Office (JDETO).

\bibliography{bibliography} 
\bibliographystyle{spiebib} 

\end{document}